\documentclass[aps,prl,showpacs,superscriptaddress, twocolumn]{revtex4-1}%
\usepackage{amsmath}
\usepackage{amssymb}
\usepackage{amsfonts}
\usepackage{float}
\usepackage{graphicx}
\usepackage[colorlinks,linkcolor=blue,anchorcolor=blue,citecolor=blue,urlcolor=blue]{hyperref}
\usepackage{mathrsfs}
\usepackage{epsfig}
\usepackage{subfigure}
\usepackage{physics}
\usepackage{bbold}
\usepackage{CJKutf8}

\begin{document}
\begin{CJK}{UTF8}{gbsn}

\title{
Emergent prethermalization signatures in out-of-time ordered correlations}
\author{Ken Xuan Wei}\thanks{These authors contributed equally to this work.}
\affiliation{Department of Physics, Massachusetts Institute of Technology, Cambridge, MA 02139}
\author{Pai Peng (彭湃)}\thanks{These authors contributed equally to this work.}
\affiliation{Department of Electrical Engineering and Computer Science, Massachusetts Institute of Technology, Cambridge, MA 02139}  
\author{Oles Shtanko}
\affiliation{Department of Physics, Massachusetts Institute of Technology, Cambridge, MA 02139}
\author{Iman Marvian}
\affiliation{Departments of Physics \& Electrical and Computer Engineering, Duke University, Durham, NC 27708}
\author{Seth Lloyd}
\affiliation{Department of Mechanical Engineering, Massachusetts Institute of Technology, Cambridge, MA 02139}
\author{Chandrasekhar Ramanathan}
\affiliation{Department of Physics and Astronomy, Dartmouth College, Hanover, NH 03755, USA} 
\author{Paola Cappellaro}\email[]{pcappell@mit.edu}
\affiliation{Department of Nuclear Science and Engineering, Massachusetts Institute of Technology, Cambridge, MA 02139}  

\begin{abstract}
How a many-body quantum system thermalizes --or fails to do so-- under its own interaction is a fundamental yet elusive concept. Here we demonstrate nuclear magnetic resonance observation of the emergence of prethermalization
by measuring out-of-time ordered correlations. We exploit Hamiltonian engineering techniques to tune the strength of spin-spin interactions and of a transverse magnetic field in a spin chain system, as well as  to invert the Hamiltonian sign to reveal out-of-time ordered correlations. At large fields, we observe an emergent conserved quantity due to prethermalization, which can be revealed by an early saturation of correlations. Our experiment not only demonstrates a new protocol to measure out-of-time ordered correlations, but also provides new insights in the study of quantum thermodynamics.
\end{abstract}

\maketitle
\end{CJK}

%
%
%
%
%
%
%
The dynamics of many-body quantum systems  can display a multitude of interesting phenomena, ranging from thermalization~\cite{Kaufman16s, kucsko18l} to many-body localization (MBL)~\cite{Basko06, Basko07, Nandkishore15, Schreiber15, Smith16, Choi16, Wei18, Lukin18x}, discrete time crystals~\cite{Khemani2016, Keyserlingk2016, Else2016, Moessner2017, Sacha2018, Yao2017, Zhang2016, Choi2017, Ho2017}, and dynamical phase transitions~\cite{Zhang2017, Yuzbashyan2006, Heyl2013, Zunkovic2018, Flaschner2016a, Jurcevic2017}. 
Recently, there has been increased interests in systems exhibiting nonergodic dynamics in the absence of any disorder or incommensurate fields, such as quasi-MBL in translationally invariant systems~\cite{Yao16} and disorder free localization~\cite{Papic15, Smith17,Michailidis2018b}. Another intriguing possibility is prethermalization, where nonintegrable quantum systems may fail to thermalize on practically accessible timescales~\cite{Berges2004, Gring2012, Else2017,Else2017a,Abanin2017, Neyenhuis2017a}, due to an emergent quasi-local integral of motion. 

Here we study thermalization and prethermalization by measuring out-of-time-ordered (OTO)  commutators~\cite{Larkin1969, Kitaev2015, Kitaev2017, Garttner2017a, Li2017, Landsman2018a}, which are powerful indicators of information scrambling, but are typically difficult to observe experimentally.
%

We exploit Hamiltonian engineering techniques to investigate the 
onset of prethermalization in a nuclear spin system in a natural crystal. We can access different regimes by manipulating the relative strengths of the dipolar interactions among spins and the transverse magnetic field. After a quench, we experimentally measure OTO commutators using multiple quantum coherence (MQC) experiments~\cite{Munowitz75, Garttner2017, Wei18} for a system initially at an effective infinite temperature. 
In the low field regime, the system is thermalizing  and the commutator  keeps increasing in the observed timescale. In the high field regime, instead, an emergent conserved quantity arises due to prethermalization and the OTO commutator involving such prethermal conserved quantity  saturates after a short time. 
We further support the interpretation of our experimental results by constructing the prethermal Hamiltonian perturbatively~\cite{Else17, Abanin2017}. We numerically observe the divergence of the perturbation series below a certain transverse field threshold, indicating the breakdown of prethermal dynamics and the onset of the thermal regime.
 	

\begin{figure*}\centering
\includegraphics[width=0.95\textwidth]{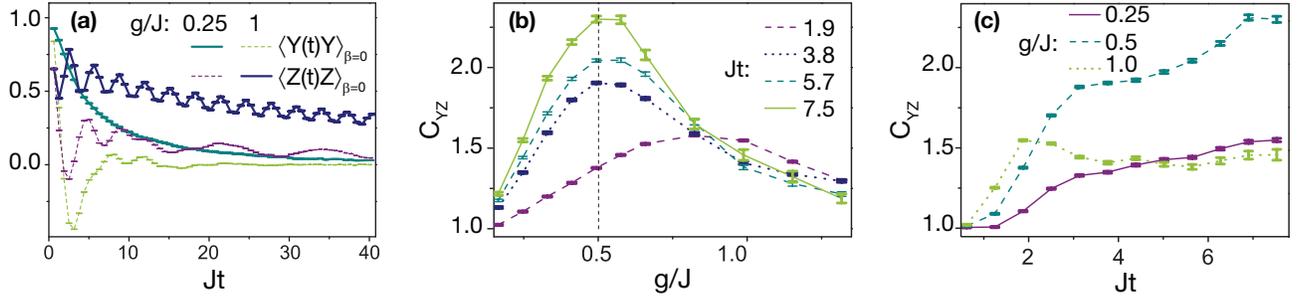}
\caption{\label{figCYZ}
(a) Distinct behavior for transverse ($Y$) and longitudinal ($Z$) magnetization:
 $\langle Y(t)Y\rangle_{\beta=0}$ at $g/J=0.25$ shows a fast decay as a function of time, indicating thermalization and erasure of initial memory. $\langle Z(t)Z\rangle_{\beta=0}$ at $g/J=1$  shows instead slow nonergodic dynamics with periodic oscillations. In the opposite regimes (dashed lines) both correlations quickly decay to zero.
Experimentally measured OTO commutator, $C_{\mathrm{YZ}}$, as a function of transverse field strength (b) and normalized time (c). We observe the fastest growth around $g/J=0.5$ (marked by a dashed line in (b)).
Here and in the rest of the paper, error bars are determined from the noise in the free induction decay 
(see Supplementary Information for details on the experimental scheme).  
}
\end{figure*}
We work with nuclear spins in fluorapatite (FAp)~\cite{VanderLugt64}, an experimental system recently used to show MBL~\cite{Wei18}. The $^{19}$F spins-1/2 form linear chains in the crystal and are coupled by the magnetic dipolar interaction. A single crystal  is placed in a large (7~T) magnetic field at room temperature. 
In a strong magnetic field the interaction Hamiltonian for the  $^{19}$F spins is given by the secular dipolar Hamiltonian  $H_\text{Dipz}=\sum_{j,k>j}J_{jk}\!\left[S_z^j S_z^k -\frac{1}{2}(S_x^j S_x^k +S_y^j S_y^k) \right]
$,
where $J_{jk}= J|j-k|^{-3}$. Here \(S_\alpha^j\) \((\alpha=x,y,z)\) are spin-1/2 operators of the \(j\)-th $^{19}$F spin 
(see Section I in SM). 
In the timescales we explore, the system can be approximately treated as an ensemble of identical spin chains~\cite{Cappellaro07l, Zhang09, Ramanathan11}, since the interchain coupling is $\sim\!40$ times weaker than the intrachain coupling. The coupling to $^{31}$P spins in the lattice is refocused by the applied control, and the spin-lattice relaxation effects are negligible ($T_1\approx0.8$~s). The dynamics of the $^{19}$F spins are thus well approximated by a 1D closed quantum system with dipolar couplings.
While the corresponding 1D, nearest-neighbor  XXZ Hamiltonian is integrable~\cite{Alcaraz1987, Sklyanin1988, Wang2016},  the Hamiltonian we consider can lead to diffusive~\cite{Sodickson95,Zhang98l} and chaotic behavior~\cite{Jyoti17x} in 3D. In the presence of a transverse field,  the system is known to show a quantum phase transition~\cite{Isidori2011}. 
 
In experiments we consider the dynamics under an engineered Floquet Hamiltonian, obtained by modulating $H_\text{Dipz}$ with periodic sequences of strong rf pulse~\cite{Haeberlen68} that can also introduce quenches and time reversal. To  lowest order of Magnus expansion the pulse sequence (see Section II in SM) engineers a dipolar Hamiltonian along the $y$ direction, $H_\text{dipy}$~\cite{Wei18}, while an effective static transverse field is introduced by  phase shifting all the pulses.  
The resulting Floquet-Trotter Hamiltonian 
is equivalent to its lowest order
to a  transverse field dipolar Hamiltonian $H_\text{TDip}\!=\!uH_\text{Dipy}\!+\!gZ$~\footnote{Note that the form of this Hamiltonian is reminiscent of a spin-locking dipolar interaction in NMR~\cite{Slichter}}:
\begin{align}
H_\text{TDip}\!=\!u\!\!\sum_{j,k>j}\!\!J_{jk}\!\left[S_y^j S_y^k \!-\!\frac{1}{2}(S_x^j S_x^k\! +\!S_z^j S_z^k) \right]\!+\!g\!\sum_j\! S_z^j
\label{eq:ham}
\end{align}
where both $u$ and $g$ 
are under experimental control (for details see Section II in SM) and we set $J=-uJ_{j,j+1}$ being the engineered nearest-neighbor coupling strength. 
In all experiments we set $u=0.2$ and pulse sequence period $t_c=96~\mu$s, which corresponds to an effective $J t_c=0.62$, given the natural $J_{j,j+1}=-33$~krad/s neighbor coupling strength in fluorapatite.
Whereas for either $g\!=\!0$ or $J\!=\!0$ the magnetizations $Y=\sum_j S_y^j$ and $Z=\sum_j S_z^j$ are exactly conserved, respectively, for finite values their dynamics is quite different, as observed in the experimental two point correlators $\langle Z(t)Z\rangle_{\beta=0}$ and $\langle Y(t)Y\rangle_{\beta=0}$ shown in Fig.~\ref{figCYZ}(a).
Indeed,  $H_\text{TDip}$ can be mapped to a prethermal Hamiltonian only at   high field, while at low field we expect thermalization. 
As shown in Ref.~\cite{Abanin17b}, a prethermal regime exists for Hamiltonians that can be divided into two parts $H\!=\!H_0+\epsilon V$, with $H_0$ having integer eigenvalues up to a scaling factor $C$, $e^{i2\pi C H_0}\!=\!\mathbb{1}$.
For sufficiently small $\epsilon$,  $H$ can be approximately transformed to a prethermal Hamiltonian $H_\mathrm{pre}$,  through a local unitary $R$~\cite{Abanin2017,Else17b}, i.e. $RHR^\dagger=H_\text{pre}+ \delta H$, where $\delta H$ is exponentially small in $\epsilon$,  $R=\mathbb{1}+O(\epsilon)$, and $[H_0',H_\text{pre}]=0$, where $H_0'$ in the frame rotated by $R$ has the same matrix representation as $H_0$ in the original frame, so they are different physical operators.
As the prethermal Hamiltonian conserves $H'_0$,   
$R^\dagger  H_0' R$ is  a conserved quantity in the original frame up to an exponentially long time $t_{\text{pre}}$, after which  the small correction $\delta H$ thermalizes the system.
In the transverse field dipolar model with $g\gg J$, we can identify the dominant part with the field and the perturbation with the dipolar interaction. The prethermal Hamiltonian is then $H_\mathrm{pre}= g Z'-u H'_\text{Dipz}/2+O(J/g)$. 
Then, in the prethermal regime we expect  an emergent conserved quantity,  $Z_\text{pre}$,  related to $Z'$ by a local unitary transformation $R$, $Z_\text{pre}= R^\dag Z' R$.

To investigate the presence of this emergent constant of motion beyond the partial information given by local observables [Fig.~\ref{figCYZ}(a)]  we experimentally analyze the properties of OTO commutators [Fig.~\ref{figCYZ}(b-c)],  defined as  $C_\mathrm{AB}(t)\equiv\langle [A(t),B][A(t),B]^\dagger\rangle_\beta$, where $A(t)=U(t)AU(t)^\dag$, with $U(t)=e^{-i\hat Ht}$ and $\hat{H}$ being the system Hamiltonian. Here $\langle \cdot\rangle_\beta=\mathrm{Tr}(e^{-\beta\hat H}\,\cdot\,)/\mathrm{Tr}(e^{-\beta\hat H})$ denotes the ensemble average at the inverse temperature $\beta$. The OTO commutator contains a term with an unconventional temporal order, the OTO correlator $F(t)\equiv\langle {A}^\dagger(t){B}^\dagger A(t) B\rangle_\beta$, which can  provide a more accurate description of operator scrambling than, e.g., Loschmidt echoes~\cite{Quan2006,Swingle16, Hahn1950, Rhim1971, Zhang1992, Andersen2003, Gorin2006, Prosen2003}. 
We  exploit our ability to engineer a time reversal of the Hamiltonian in Eq.~(\ref{eq:ham}) to measure the OTO commutator of extensive observables~\cite{Kukuljan17}, as we explain in the following.

In room temperature NMR experiments, the initial state for a chain of $L$ spins is described by the density matrix $\rho(0)\!\approx\!(\mathbb{1}\!-\!\epsilon Z)/2^L$, with  $\epsilon\! \sim\! 10^{-5}$. Since the identity operator does not contribute to any measurable signal, we only care about the deviation from it, $\delta\rho=2Z/\sqrt{L}$, which has been normalized such that $\text{Tr}(\delta\rho^2)/2^{L}=1$. The mixed initial state enables the experimental study of two-point correlators and OTO commutators in a straightforward way. Since $\delta\rho(0)$ is usually the collective spin magnetization pointing in some direction, $\mathcal{O}_{\bf n}=\sum_j {\bf n} \cdot {\boldsymbol S}_j$ and we can measure the collective magnetization around any axis, the typical signal is the two-point correlation, $4\text{Tr}[U(t)\delta\rho(0)U^{\dag}(t)\mathcal{O}_{\bf n}]/(2^{L}L)\equiv \langle \mathcal{O}_{\bf n}(t)\mathcal{O}_{\bf n}\rangle_{\beta=0}$. That is, in our experiments, the (deviation of) the density matrix plays the role of an observable for an effective simulated system at infinite temperature. Crucially, however, the ``simulated observable'' $\delta\rho$ will thermalize at long times  under the strong driving, $\langle\delta\rho(t)\rangle=0$: this enables distinguishing the prethermal regimes from the expected (zero) signal at long times due to the eventual thermalization. 
MQC experiments~\cite{Munowitz75,Baum85,Ramanathan03}  measure the overlap of the time-evolved density matrix, $\delta\rho(t)=U(t)\delta\rho(0)U^{\dag}(t)$, with itself after a collective rotation. The overall measured signal can be expressed as 
\begin{align}
S_\phi=
2^{-L}\text{Tr}[ e^{-i \phi \mathcal{O}_{\bf n}}\delta\rho(t) e^{i \phi \mathcal{O}_{\bf n}}\delta\rho(t)].
\label{eq: mqc_sig}
\end{align}
Taking a discrete Fourier transform of $S_\phi$ with respect to $\phi$ yields the MQC intensities: $S_\phi=\sum_q e^{-i q \phi}I_q$.
Expanding $S_\phi$ in powers of $\phi$, it can be shown that $\text{Tr}( [\delta\rho(t), \mathcal{O}_{\bf n}] ^2)/2^{L}\!=\!-\sum_q q^2 I_q$.  Setting  $\delta\rho(0)=\mathcal{O}_{\bf n'}$, we can write
\begin{align}
C_{\mathcal{O}_{\bf n'}\mathcal{O}_{{\bf n}}}(t)=\frac4L\langle |[\mathcal{O}_{\bf n'}(t), \mathcal{O}_{\bf n}]|^2\rangle_{\beta=0}=\sum_q q^2 I_q(t)
\label{eq:com}
\end{align}
Eq.~(\ref{eq:com}) is the central idea of our experiments: by measuring the second moment of the MQC intensities encoded in $\delta\rho(t)$ along $\mathcal{O}_{\bf n}$ one can obtain the OTO commutator  between $\mathcal{O}_{\bf n'}(t)$ and $\mathcal{O}_{\bf n}$ as if the system were at infinite temperature~\footnote{Notice that exchanging $\mathcal{O}_{\bf n'}$ and $\mathcal{O}_{\bf n}$ will result in a different MQC distribution $I_q$, however its second moment remains the same.}. Eq.~(\ref{eq:com}) was first derived in a different context in Ref.~\cite{Khitrin1997cpl} for NMR systems. When applied to pure states,
 it relates the second moment of the MQC distribution to the quantum Fisher information~\cite{Garttner2017}. 

To study the system dynamics after a quench to Hamiltonian~(\ref{eq:ham}), we measure the OTO commutator $C_{\mathrm{YZ}}\equiv 4L^{-1}\langle |[Y(t),Z]|^2\rangle_{\beta=0}$  for various transverse field strengths and times [see Fig.~\ref{figCYZ}(b-c)]. 
First note that in the limit $g\rightarrow \infty$, $Z$ is a conserved quantity thus making $C_\mathrm{YZ}$ constant. 
In Fig.~\ref{figCYZ}(c) we observe that for large but finite transverse field $C_\mathrm{YZ}$ stops growing at an early time, revealing that $Z$ is approaching the emergent conserved quantity $Z_\text{pre}$ -  as also indicated by the slow decay and persistent oscillation of the two point correlator $\langle Z (t)Z \rangle_{\beta=0}$ [Fig.~\ref{figCYZ}(a)]. 
For small transverse field, instead, $C_{\mathrm{YZ}}$ keeps increasing, suggesting that the system is thermalizing  [Fig.~\ref{figCYZ}(c)]. We note that in the limit of exactly no transverse field, $Y$ is a conserved quantity thus making $C_{\mathrm{YZ}}$ constant. However, as long as a small field is introduced  the system becomes thermal until the field strength induces a transition to a prethermal state: we thus observe a maximum of $C_\mathrm{YZ}$ at around $g/J \approx 0.5$  [Fig.~\ref{figCYZ}(b)]. The thermal dynamics for an initial effective infinite temperature state is further  indicated by the decay of $\langle Y(t)Y\rangle_{\beta=0}$ in Fig.~\ref{figCYZ}(a)  and  additional OTO commutators presented below.
%
%

\begin{figure}[t]\centering
\includegraphics[width=0.45\textwidth]{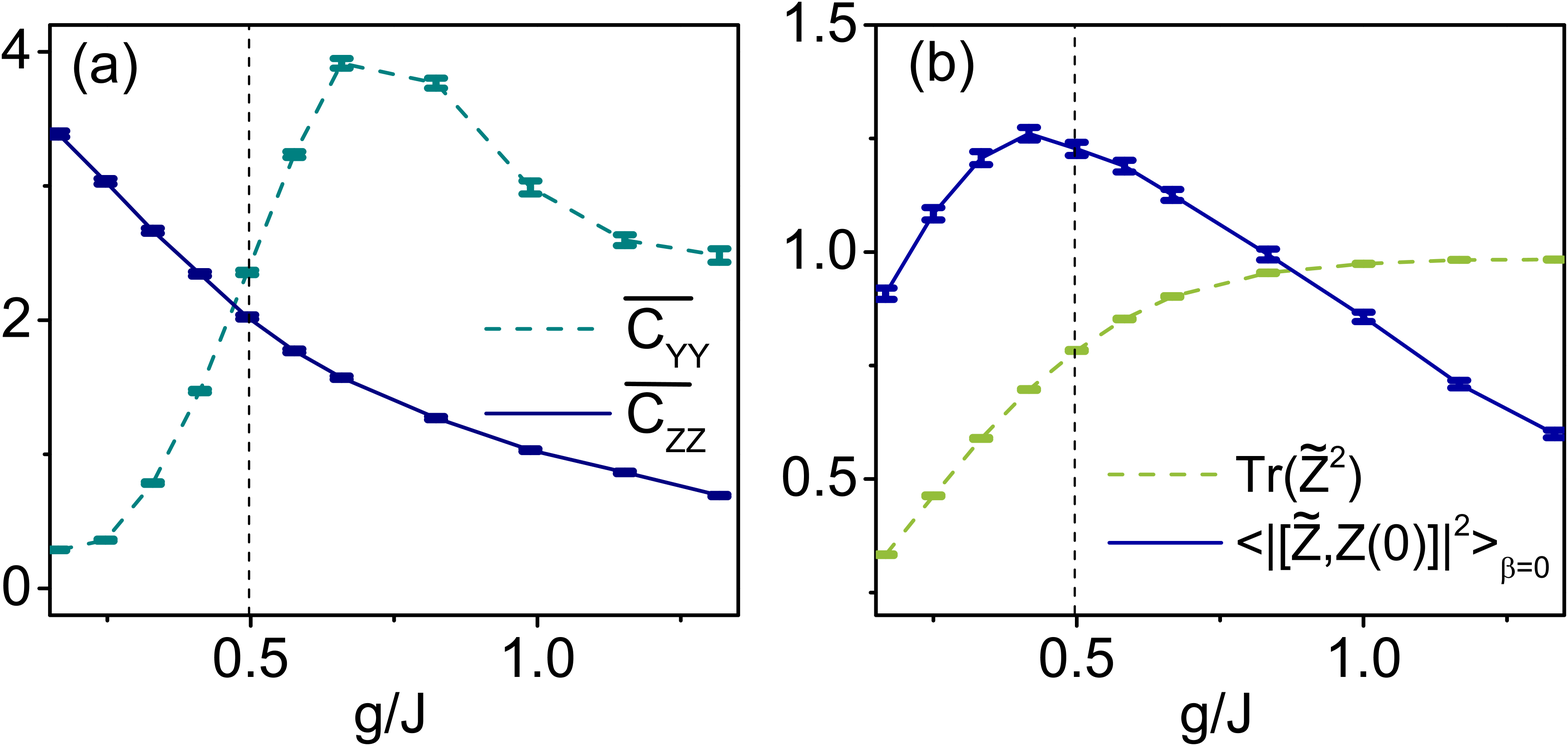}
\caption{\label{figAVE}
(a) Averaged $C_{\mathrm{YY}}$ (dashed) and $C_{\mathrm{ZZ}}$ (solid) as a function of  transverse field strength.
(b)  $\mathrm{Tr}(\widetilde{Z}^2)$ (dashed) and $\langle|[\widetilde{Z},Z(0)]|^2\rangle_{\beta=0}$ (solid) versus transverse field strength. 
The time average is taken over the values $Jt=3.77, 5.02, 6.28, 7.54, 8.80, 10.05$, with the longest time corresponding to 16 cycles ($1.54$ms).
}
\end{figure}

To gain further insight into the differences between the thermalizing and  prethermal regimes, we measure  $C_{\mathrm{ZZ}}$ and $C_{\mathrm{YY}}$, as shown in Fig.~\ref{figAVE}(a). Because these OTO commutators fluctuate significantly in time, we average them at six different times. 
As $g$ increases, $Z(t)$ approaches the prethermal conserved quantity $Z_\text{pre}$, which itself gets close to $Z$, and $C_{\mathrm{ZZ}}$ gets smaller. 
This behavior is only observed for OTO commutators involving at least one operator that overlaps with the  emergent conserved quantity, while other commutators, such as $C_{\mathrm{YY}}$, keep growing as if the system were thermal, regardless of the transverse field strength (with the exception of exactly zero field, $g=0$).

While we cannot directly measure $Z_\mathrm{pre}$,  the time-averaged operator $\overline Z= t^{-1}_\text{pre}\int_0^{t_{\text{pre}}} Z(t) dt$ (where $t_\text{pre}$ is the timescale over which the prethermal conserved quantity is present) captures its essential features~\footnote{In the limit $t_\text{pre}\!\!\rightarrow\!\infty$, $\overline Z$ becomes an exact integral of motion and it has also recently been proposed as an experimentally accessible observable for detecting MBL to ergodic phase transition~\cite{Chandran2015b}}. Indeed, we can generally write $Z(t)=Z_\text{pre} + U(t)(Z-Z_\mathrm{pre})U(t)^\dag$: then,  in the prethermal regime, the second term is small and fluctuates,  yielding  $\overline Z\approx Z_\text{pre}$ after time average.
We can approximate $\overline Z$ with a discrete time average, $\widetilde{Z}=\sum_{n=1}^{N}  Z(t_n)/N$, by independently varying the forward and backward evolution times in the MQC protocol  (see SM for details on the experiments and for a comparison between $\overline Z$ and $\widetilde Z$).
Figure~\ref{figAVE}(b) shows that $\mathrm{Tr}({\widetilde{Z}}^2)/\mathrm{Tr}({Z(0)}^2)\to1$ as $g$ increases, because the time-varying part of $Z(t)$ is very small for large $g$. 
Furthermore, $4/ L\langle|[\widetilde{Z},Z(0)]|^2\rangle_{\beta=0}$ approaches zero at large $g$, suggesting that $\lim_{g\to\infty} \widetilde Z= Z$. 

To support our interpretation of the experimental results, we explicitly construct the prethermal Hamiltonian, showing that indeed $Z_\text{pre}\approx Z$ is an emergent constant of motion. 
 \begin{figure*}\centering
\includegraphics[width=0.98\textwidth]{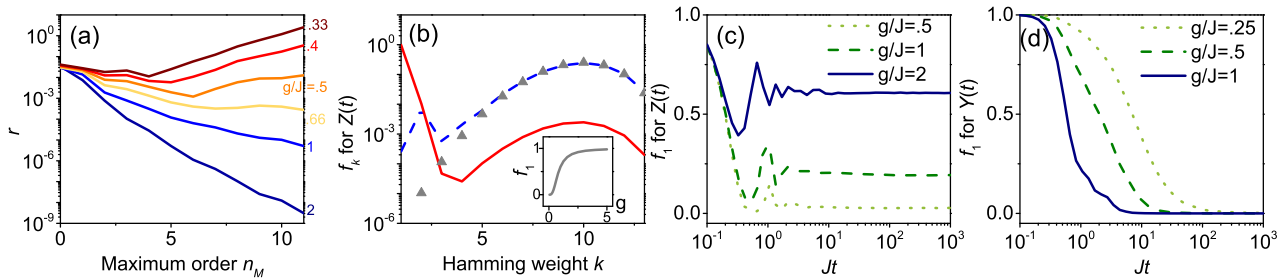}
\caption{\label{fig:num}
(a) Eigenvalue difference $r$ with respect to maximum order $n_M$ for different values of $g/J$. 
$r$ shows a divergence, up to a maximum field value $g/J=0.5$.
(b) Decomposition of $Z(t)$ (obtained by exact diagonalization) at $Jt=10^3$ 
according to the operator Hamming weight: $f_k$ is the contribution of all possible spin correlations with Hamming weight $k$. 
For small fields, $g/J=0.05$ (dashed line), the result follows closely the distribution (triangles) obtained  randomly sampling all possible operators. For large fields, $g/J=5$ (solid line) there is a significant contribution of single-body terms, related to the quasi-conserved quantity $Z_\text{pre}$. In the inset:  $f_1$ as a function of $g$.
$f_1$ for $Z(t)$ (c) and $Y(t)$ (d) as a function of normalized time, showing the nonthermal behavior of $Z$ at large $g/J$, while $Y$ is always thermal even for small $g/J$.
The system size is $L=12$ for (a) and 13 for (b--d).
}
\end{figure*}
The prethermal Hamiltonian can be expanded in powers of $\epsilon=J/g$
\begin{equation}
H_\mathrm{pre}=Z'+\sum_{n=1}^{n_M} \epsilon^n h^{(n)},
\label{eq:hpre}
\end{equation}
and numerically evaluated up to order $n_M$ (see Section IV in SM). 
It has been shown~\cite{Abanin2017} that for generic many-body systems the series in Eq.~(\ref{eq:hpre}) might not converge as $n_M\rightarrow \infty$, but there  exists an optimal order $n^*$ when truncating the series, so that $H_\text{pre}$ is most similar to $H$. If the system Hamiltonian does indeed support a prethermal phase, we expect its eigenvalues $E_m$ to be close to the prethermal Hamiltonian ones, $E_m^\mathrm{pre}$. We thus calculate  the eigenvalue difference $r\equiv \mathrm{mean}_m (E_m-E_m^\mathrm{pre})/L$ (where $m$ labels the eigenvalues in ascending order), expecting $r$ to converge to zero only in the prethermal phase. 
Figure~\ref{fig:num}(a) shows $r$ as a function of  maximum truncation order $n_M$ for different values of $\epsilon$. 
For large $g$, $r\approx0$ appears to converge up to the largest numerically accessible order, suggesting that $H_\mathrm{pre}$ is similar to $H$ and there exists an approximately conserved quantity $Z_\mathrm{pre}$. For small $g$ however, $r$  diverges,  indicating that a prethermal Hamiltonian that conserves $Z'$ cannot be found. The transition happens at around $g/J=0.5$. 

To further demonstrate that a conserved quantity emerges for large $g$, we simulate $Z$ at large times $(Jt=10^3)$ and decompose it according to the Hamming weight~\cite{Wei18}
\begin{equation}
Z(Jt=10^3)=\sqrt{2^{L-2}L}\sum_{k=1}^L\sum_{s=1}^{\zeta_k}b_k^s(Jt)\mathcal{B}_k^s,
\end{equation}
where $\mathcal{B}_k^s$ are operators composed of tensor products of
$k$ Pauli matrices and $L-k$ identity operators, and $\zeta_k\propto 3^k \times {L\choose k}$
labels the number of configurations with  $k$ non-identity Pauli operators. 
We define the  Hamming weight of $k$-spin correlations as $f_k=\sum_{s=1}^{\zeta_k} [b_k^s]^2$, satisfying $\sum_{k=1}^L f_k=1$. Figure~\ref{fig:num}(b) shows that for  small transverse field $f_k$ is approximately proportional to $\zeta_k$, suggesting that all possible operators $\mathcal{B}_k^s$ have the same weight, in agreement with the eigenvalue thermalization hypothesis~\cite{Hosur2016,Rigol08,Deutsch1991a,Srednicki1994e}. 
The result is qualitatively different for $g\gg 1$, where a significant one-body term, $f_1$, exists even at very late times ($Jt=10^3$), signifying the failure of thermalization and the emergence of the conserved quantity $Z_\text{pre}$.
We thus study $f_1$ as a function of time. For small fields, $g/J\leq0.5$, the contribution of $f_1$ in $Z(t)$ quickly relaxes from one to zero, as shown in Fig.~\ref{fig:num}(c). 
For  large $g/J$, instead, $f_1$ reaches  a non-zero, quasi-equilibrium value, signaling the prethermal phase. We do not see the final thermalizing stage in the numerics, possibly because  small systems, $L=13$, do not fully thermalize~\cite{Machado2017}. 
On the other hand, while $Y$ is conserved at exactly zero field ($g=0$), as soon as a small transverse field is introduced the contribution of $f_1$ to $Y(t)$ decays to zero  [Fig.~\ref{fig:num}(d)]. 
This indicates that the slow dynamics observed for $Y(t)$ at small $g$ is not protected by any prethermal conserved quantity, and will thus thermalize on timescales much shorter than $t_\text{pre}$. 
The quantitative difference between $f_1$ for $Y(t)$ and  $Z(t)$  can be approximately observed by measuring the two-point correlations $\langle Z(t)Z\rangle_{\beta=0}$ and $\langle Y(t)Y\rangle_{\beta=0}$.
As shown in Fig.~\ref{figCYZ}(a), in the small field regime $\langle Y(t)Y\rangle$ decreases rapidly as a function of time, suggesting that $f_1(t)$ of $Y(t)$ is not a (prethermal) conserved quantity~\footnote{We further numerically demonstrate in the SM that $H$ is unlikely to be in a prethermal regime with respect to NN Ising interactions $\sum_j S_y^j S_y^{j+1}$, the dominant term in $H_\text{TDip}$ for small values of $g$.}. In stark contrast, $\langle Z(t)Z\rangle$ shows a slow decay with periodic oscillations, suggesting that $f_1(t)$ of $Z(t)$ is mostly conserved, consistent with prethermalization at large $g/J$~\footnote{taking the average over time, also shows that $\overline Z\approx Z$ in the prethermal regime}. 

In conclusion, we studied the out-of-equilibrium dynamics of the transverse field dipolar interaction in a solid-state NMR quantum simulator. Using MQC techniques, we measured OTO commutators to reveal a distinct dynamics in the high and low field regimes, and identified them as prethermal and thermal phases. In the prethermal regime, when one of the OTO operators is close to the emergent quasi-conserved quantity, the OTO commutator saturates at an early time, while it keeps increasing in the thermal regime, with a transition at about $g/J=0.5$~\footnote{we note that this value does not corresponds to the critical point of the equilibrium quantum phase transition~\cite{Isidori2011}}. 
We further validate our experimental results numerically, by constructing the  prethermal Hamiltonian and  verifying the emergence of a conserved quantity at high field, while in the low field regime the dynamics is consistent with thermalization.   
We demonstrate the value of OTO commutators in  investigating  non-equilibrium quantum thermodynamics, while also providing a method to experimentally measure OTO commutators that could be extended to other experimental platforms. Similar techniques could be used for example to explore other many-body phenomena, such as localization, dynamics phase transition and information scrambling, paving the way to more comprehensive understanding of out-of-equilibrium quantum many-body systems.  

We thank N. Halpern, D. Huse, and I. Cirac for insightful discussions. This work was supported by the National Science Foundation PHY1734011.

\bibliography{library1,Biblio}

\newpage
\onecolumngrid
\newpage

\section*{SUPPLEMENTARY MATERIAL}
%
%
\section{I. Experimental System}
The system used in the experiment was a single crystal of fluorapatite (FAp).  Fluorapatite is a hexagonal mineral with space group \(P6_3/m\), with the \(^{19}\)F spin-1/2 nuclei  forming linear chains along the \(c\)-axis. Each fluorine spin in the chain is surrounded by three \(^{31}\)P spin-1/2 nuclei.
We used a natural crystal, from which we cut a sample of approximate dimensions 3 mm$\times$3 mm$\times$2 mm.
The sample is placed at room temperature inside an NMR superconducting magnet producing a uniform $B=7$ T field. The total Hamiltonian of the system is given by
\begin{equation}
H_{tot}=\omega_F \sum_k S_z^k+\omega_P \sum_\kappa s_z^\kappa+H_{F}+H_P+H_{FP}
\label{eq:Hamtot}	
\end{equation}
The first two terms represent the Zeeman interactions of the F($S$) and P($s$) spins, respectively, with frequencies $\omega_F=\gamma_FB\approx (2\pi)282.37$ MHz and $\omega_P=\gamma_PB=(2\pi)121.51$ MHz, where $\gamma_{F/P}$ are the gyromagnetic ratios. The other three terms represent the natural magnetic dipole-dipole interaction among the spins, given generally by
\begin{equation}
  H_{dip}=\sum_{j<k}\frac{\hbar\gamma_j\gamma_k}{|\vec r_{jk}|^3}\left[\vec S_j\cdot\vec S_k-\frac{3\vec S_j\cdot\vec r_{jk}\,\vec S_k\cdot\vec r_{jk}}{|\vec r_{jk}|^2}\right],
\end{equation}
where $\vec r_{ij}$ is the vector between the $ij$ spin pair. Because of the much larger Zeeman interaction, we can truncate the dipolar Hamiltonian to its energy-conserving part (secular Hamiltonian). We then obtain the homonuclear Hamiltonians
\begin{equation}
  H_F=\frac{1}{2}\sum_{j<k}J^F_{jk}(2 S_z^j S_z^{k}- S_x^j S_x^{k}- S_y^j S_y^{k}) \qquad H_P=\frac{1}{2}\sum_{\lambda<\kappa}J^P_{\kappa\lambda}(2s_z^\lambda s_z^{\kappa}-s_x^\lambda s_x^{\kappa}-s_y^\lambda s_y^{\kappa})
\end{equation}
and the heteronuclear interaction between the $F$ and $P$ spins,
\begin{equation}
  H_{FP}=\sum_{k,\kappa} J^{FP}_{k,\kappa}S_z^ks_z^\kappa,
\end{equation}
with $J_{jk}=\hbar\gamma_j\gamma_k\frac{1-3\cos(\theta_{jk})^2}{|\vec r_{jk}|^3}$, where $\theta_{jk}$ is the angle between the vector $\vec r_{jk}$ and the magnetic field $z$-axis. The maximum values of the couplings (for the closest spins) are given respectively by $J^F=-32.76$ krad s$^{-1}$, $J^P=1.20$ krad s$^{-1}$ and $J^{FP}=6.12$ krad s$^{-1}$. 
\begin{figure}[b]
\centering 
\includegraphics [width=0.8\linewidth ]{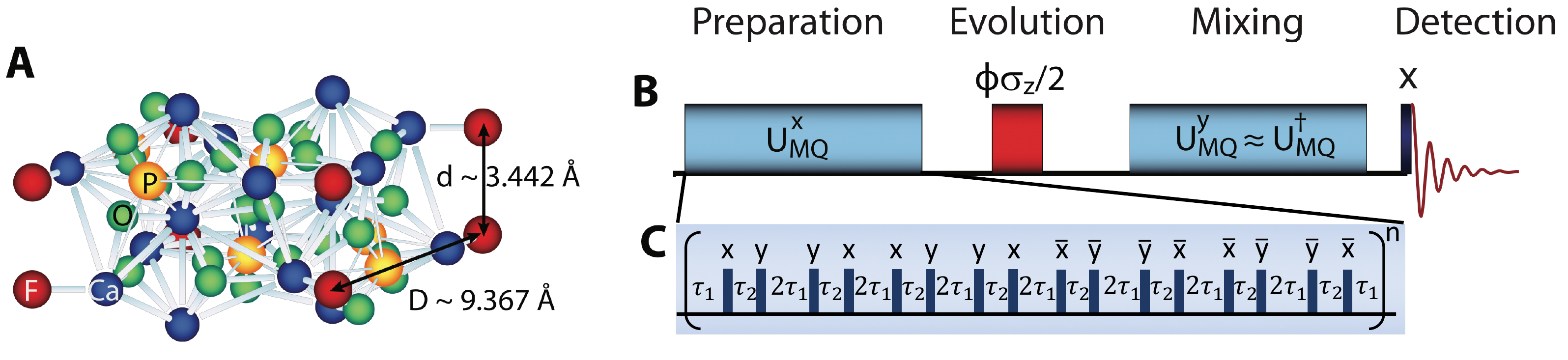}
\caption{\textbf{A} Fluorapatite crystal structure, showing the Fluorine and Phosphorus spins in the unit cell. \textbf{B} NMR scheme for the generation and detection of MQC. In the inset  (\textbf{C}) an exemplary pulse sequence for the generation of the $H_\mathrm{dipy}$. Note that thanks to the ability of inverting the sign of the Hamiltonian, the scheme amounts to measuring out-of-time order correlations.
}\label{fig:mqcd}
\end {figure}

The dynamics of this complex many-body system can be  mapped to a much simpler, quasi-1D system. First, we note that when the crystal is oriented with its $c$-axis parallel to the external magnetic field
the coupling of fluorine spins to the closest off-chain fluorine spin is $\approx40$ times weaker, while in-chain, next-nearest neighbor couplings are $8$ times weaker. 
 Previous studies on these crystals have indeed observed dynamics consistent with spin chain models, and the system has been proposed as solid-state realizations of quantum wires
~\cite{Cappellaro07l,Cappellaro11,Ramanathan11}. This approximation of the experimental system to a 1D, short-range system, although not perfect has been shown to reliably describe experiments for relevant time-scales~\cite{Rufeil-Fiori09b,Zhang09}. The approximation breaks down at longer time, with a convergence of various effects: long-range in-chain and cross chain couplings, as well as pulse errors in the sequences used for Hamiltonian engineering. In addition, the system also undergoes spin relaxation, although on a much longer time-scale ($T_1=0.8~$s for our sample). 

\subsection{Error analysis}
\begin{figure}[h]
\centering
\includegraphics[width=90mm,clip]{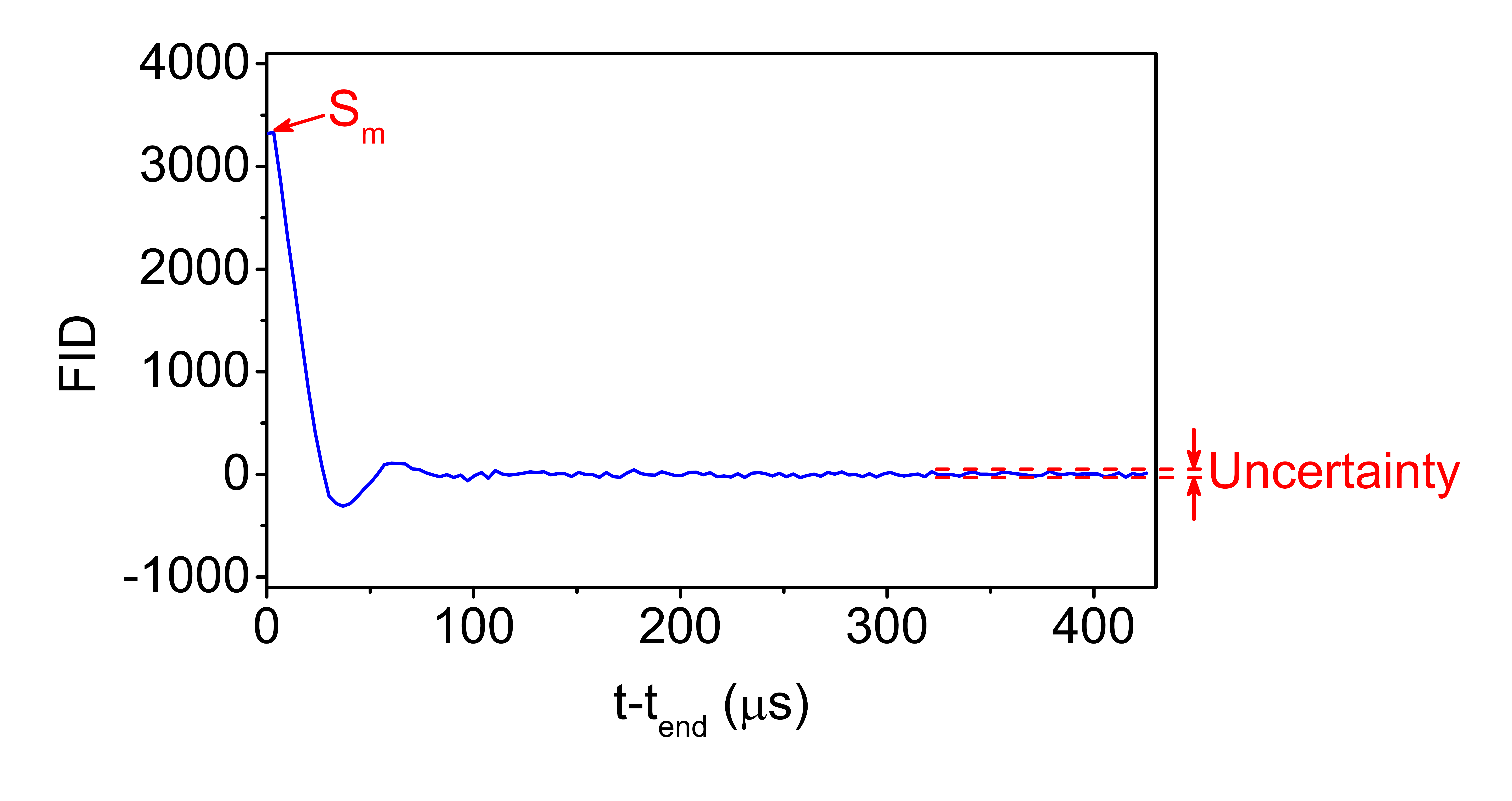}
\caption{\label{fig:FID}
An example of FID. 128 data points are taken in total. The first data point gives $S_m$ and the standard deviation of the last 20 points gives the uncertainty of $S_m$.
}
\end{figure}

The desired quantity $S_m=\mathrm{Tr}(\delta\rho(t_\mathrm{end})\mathcal{O})$, where $\delta\rho(t_\mathrm{end})=U_{-t}e^{-i \frac{m \pi P}{L}} U_t\delta\rho(0)U_{-t} e^{i \frac{m\pi P}{L}} U_t$ is the nontrivial part of the density matrix at the end of the entire pulse sequence $t_\mathrm{end}$. To get the uncertainty of $S_m$, instead of just measuring one point, we continuously monitor the free evolution of $\delta\rho(t_\mathrm{end})$ under the natural Hamiltonian $H_\mathrm{dipz}$, so called free induction decay (FID, a typical FID process is shown in Fig. \ref{fig:FID}). We take the standard deviation of the last 20 data points in the FID as the uncertainty of the $S_m$, and then linearly propagate to  get the error bar of the OTO commutators.

\section{II. Hamiltonian Engineering}
The Hamiltonians used in the main text were obtained stroboscopically (Floquet Hamiltonians) by applying periodic rf pulse trains to the natural dipolar Hamiltonian that describes the system. We used Average Hamiltonian Theory (AHT~\cite{Haeberlen68}) as the basis for our Hamiltonian engineering method, to design the control sequences and determine the approximation errors.

To see how repeatedly applying a periodic pulse sequence modifies the dynamics of the system, we write the total Hamiltonian as \(H=H_\text{dip}+H_\text{rf}\), 
where \(H_\text{dip}=\frac{1}{2}\sum_{j<k}J_{jk}(2 S_z^j S_z^{k}-S_x^j S_x^{k}-S_y^j S_y^{k})+\sum_j h_j S_z^j\) is the system Hamiltonian, 
and \(H_\text{rf}(t)\) is the external Hamiltonian due to the rf-pulses. 
The density matrix \(\rho\) evolves under the total Hamiltonian according to \(\dot\rho=-i[H,\rho]\). 
Consider an interaction frame defined by \(\rho'={U_\text{rf}}^{\dagger}\rho U_\text{rf}\), where \(U_\text{rf}(t)=\mathcal{T}\exp[-i\int_0^t H_\text{rf}(t') dt']\) and \(\mathcal{T}\) is the time ordering operator. 
In this \textit{toggling} frame, \(\rho'\) evolves according to \(\dot{\rho}'=-i[H(t),\rho']\), where \(H(t)={U_\text{rf}}^{\dagger}H_\text{dip} U_\text{rf}\). 
Since \(U_\text{rf}\) is periodic, \(H(t)\) is also periodic with the same period \(t_c\). 
The evolution operator over one period can be written as \(U(t_c)=\exp[-i H_\text{Flq} t_c]\), where \(H_\text{Flq}\) is called the \textit{Floquet} Hamiltonian (or in the language of NMR the \textit{Average Hamiltonian}). 
Note that if the pulse sequence satisfies the condition \(U_\text{rf}(t_c)=1\), the dynamics of \(\rho\) and \(\rho'\) are identical when the system is viewed stroboscopically, i.e., at integer multiples of \(t_c\). 
The system evolves as if  under a time-independent Hamiltonian \(H_\text{Flq}\). 
To calculate \(H_\text{Flq}\) we employ the Magnus expansion as is usual in AHT: \(H_\text{Flq}=H^{(0)}+H^{(1)}+\cdots\). 
The first two terms are given by
\begin{align*}
H^{(0)}=\frac{1}{t_c}\int_0^{t_c}H(t)dt ,\quad H^{(1)}=\frac{-i}{2t_c}\int_0^{t_c}dt_2\int_0^{t_2}dt_1[H(t_2),H(t_1)].
\end{align*}
The zeroth order of the average Hamiltonian \(H^{(0)}\) is often a good approximation to the Floquet Hamiltonian \(H_\text{Flq}\), as the first order can be set to zero by simple symmetrization of the pulse sequence. 

The basic building block of the pulse sequences we used in this work is given by a 4-pulse sequence~\cite{Kaur12,Yen83} originally developed to study MQC.
We denote a generic 4-pulse sequence as \(P(\tau_1,{\bf n}_1,\tau_2,{\bf n}_2,\tau_3,{\bf n}_3,\tau_4,{\bf n}_4,\tau_5)\), where \({\bf n}_j\) represents the direction of the \(j\)-th \(\pi/2\) pulse, and \(\tau_j\)'s the delays interleaving the pulses. In our experiments, the \(\pi/2\) pulses have a width \(t_w\) of typically 1 \(\mu\)s. \(\tau_j\) starts and/or ends at the midpoints of the pulses (see also Fig.~\ref{fig:mqcd}). In this notation, our forward 16-pulse sequence can be expressed as
\begin{gather*}
P(\tau_1,{\bf x},\tau_2,{\bf y},2\tau_1,{\bf y},\tau_2,{\bf x},\tau_1)P(\tau_1,{\bf x},\tau_2,{\bf y},2\tau_1,{\bf y},\tau_2,{\bf x},\tau_1)P(\tau_1,{\bf \bar{x}},\tau_2,{\bf \bar{y}},2\tau_1,{\bf \bar{y}},\tau_2,{\bf \bar{x}},\tau_1)P(\tau_1,{\bf \bar{x}},\tau_2,{\bf \bar{y}},2\tau_1,{\bf \bar{y}},\tau_2,{\bf \bar{x}},\tau_1)
\end{gather*}
and the backward sequence as
\begin{gather*}
P(\tau_3,{\bf y},\tau_3,{\bf x},2\tau_4,{\bf x},\tau_3,{\bf y},\tau_3)P(\tau_3,{\bf y},\tau_3,{\bf x},2\tau_4,{\bf x},\tau_3,{\bf y},\tau_3)P(\tau_3,{\bf \bar{y}},\tau_3,{\bf \bar{x}},2\tau_4,{\bf \bar{x}},\tau_3,{\bf \bar{y}},\tau_3)P(\tau_3,{\bf \bar{y}},\tau_3,{\bf \bar{x}},2\tau_4,{\bf \bar{x}},\tau_3,{\bf \bar{y}},\tau_3)
\end{gather*}
where \(\{{\bf \bar{x}},{\bf \bar{y}}\}\equiv \{{\bf -x},{\bf -y}\}\). The delays are given by
\begin{gather*}
\tau_1=\tau(1-u), \quad
\tau_2=\tau(1+2u), \quad
\tau_3=\tau(1+u), \quad
\tau_4=\tau(1-2u)
\end{gather*}
where \(\tau\) is typically 4 \(\mu\)s. The cycle time \(t_c\), defined as the total time of the sequence, is given by \(t_c=24\tau\). \(u\) is a dimensionless adjustable parameter, and is restricted such that none of the inter-pulse spacings becomes negative. We set $u=0.2$ in experiments. 

Our forward and backward pulse sequences with finite pulse width produce average dipolar Hamiltonians with opposite sign:
\begin{align*}
H^{(0)}=\pm u\sum_{j<k}J_{jk}\left[S_y^j S_y^k - \frac{1}{2}(S_x^j S_x^k +S_z^j S_z^k)\right],
\end{align*}
and \(H^{(1)}=0\) (the first order can always be set to zero by a proper symmetrization of the sequence). 
The interaction between $F$ and $P$ spins averages to zero under this pulse sequence. The average Hamiltonian corresponding to the forward sequence is presented as Eq.(\ref{eq:ham}) in the main text. 

A uniform transverse field can be introduced in $H^{(0)}$
by phase-shifting the entire pulse sequence. Consider rotating the \(n\)-th cycle of the pulse sequence by \((n-1)\phi\) around the \({\bf z}\) axis, which can be accomplished by phase shifting all the pulse directions ${\bf n}_j$ in the \(n\)-th cycle by \((n-1)\phi\). The evolution operator for each cycle is given by
\begin{align*}
U_1=e^{-i H^{(0)} t_c}, \quad 
U_2=e^{-i\phi Z}e^{-i H_\mathrm{Flq} t_c}e^{i\phi Z}, \quad
U_3=e^{-2i\phi Z}e^{-i H_\mathrm{Flq} t_c}e^{2i\phi Z}, \quad \cdots  \quad
U_n =e^{-i(n-1)\phi Z}e^{-i H_\mathrm{Flq} t_c}e^{i(n-1)\phi Z}
\end{align*}
where \(Z=\sum_j S_z^j\). The total evolution operator over \(n\) cycles is given by the product:
\begin{align*}
U(n t_c)&=U_n U_{n-1}\cdots U_3 U_2 U_1=e^{-i n\phi Z}\left[e^{i \phi Z} e^{-i H_\mathrm{Flq} t_c} \right]^n \approx e^{-i n\phi Z}e^{ -i \left(H_\mathrm{Flq}-\frac{\phi}{t_c}Z\right)nt_c}=e^{-i n\phi Z}e^{-i H_\text{t} n t_c},
\end{align*}
where the total Hamiltonian is given by \(H_\text{t}=H_\mathrm{Flq}+g Z\), with \(g=-\phi/t_c\). 
We note that for the values we considered, the approximation $\left[e^{i \phi Z} e^{-i H_\mathrm{Flq} t_c} \right]^n \approx e^{ -i \left(H^{(0)}-\frac{\phi}{t_c}Z\right)nt_c}$ is extremely good, with the error being on  order $g/J\sim 1.5\times 10^{-3}$.
The rotation approach also generates an extra term \(e^{-i n \phi Z}\), this term can be canceled in MQC experiments by rotating the encoding pulse by \(n\phi\). In the backward sequence, the phase shifting is done in reverse order. We note that the phase shift is typically not periodic over the course of the experiment (for example, for small values of $g$, even the maximum $n\phi<2\pi$). 

We note that our methods can be applied more broadly to engineer desired Hamiltonians $ H_{des}$ using only collective rotations of the spins applied to the naturally occurring Hamiltonian, $H_{nat}$. The engineered Hamiltonian is obtained by piece-wise constant evolution under rotated versions of the natural Hamiltonian under the condition
\(
\sum_k R_k H_{nat} R_k^\dag = H_{des},
\)
where $R_k$ are collective rotations of all the spins, which achieves the desired operator to first order in a Magnus expansion. Symmetrization of the sequence can further cancel out the lowest order correction. 
Using only collective pulses limits which Hamiltonians  can be engineered, due to symmetries of the natural Hamiltonian and the action of collective operators. For typical two-body interactions of spin-1/2, an efficient tool to predict which Hamiltonian are accessible is to use spherical tensors~\cite{Ajoy13l}.

\newpage
\section{III. Multiple Quantum Coherence}
Multiple quantum coherence (MQC) can be understood as a Fourier decomposition of a many-spin operator $\mathcal{O}$ with respect to another operator $\mathcal{P}$ satisfying $e^{-2i m \pi P}=\mathbb{1}$, where $m$ is an integer. For simplicity consider the case where $P=\sum_j {\bf S}_j\cdot {\bf n}$ is the generator of global spin rotation around the ${\bf n}$-axis. Any traceless hermitian operator $\mathcal{O}$ can be decomposed as
\begin{align}\label{eq:OOq}
\mathcal{O}=\sum_{q=-L}^{q=L} \mathcal{O}_q
\end{align}
where $L$ is the total number of spins in the system, and $\mathcal{O}_q$ satisfies following two equivalent relations:
\begin{align}\label{eq:OqP}
e^{-i \theta P} \mathcal{O}_q e^{i \theta P} =e^{-i q \theta} \mathcal{O}_q, \qquad [P,\mathcal{O}_q]=q \mathcal{O}_q.
\end{align}
From the first relation it is evident that $\mathcal{O}_{-q} = \mathcal{O}_{q}^\dagger$. 
For a given $\mathcal{O}$, each multiple quantum component $\mathcal{O}_q$ can be formally written as
\begin{align}\label{eq:mqcdef0}
\mathcal{O}_q=\frac{1}{2\pi}\int_0^{2\pi}d\theta e^{i q \theta}e^{-i \theta P} \mathcal{O} e^{i \theta P}.
\end{align}
which indicates that the multiple quantum components $\mathcal{O}_q$ come from the Fourier decomposition of $\mathcal{O}$ with respect to $P$. In addition, it can also be shown that different $\mathcal{O}_q$ are orthogonal: $\text{Tr}(\mathcal{O}_q \mathcal{O}_{q'})\propto \delta_{q(-q')}$.
Using the fact that the maximum $q$ allowed is equal to the total number of spins $L$, one can replace Eq. (\ref{eq:mqcdef0}) as a discrete Fourier transform:
\begin{align}
\mathcal{O}_q=\frac{1}{2L}\sum_{m=0}^{2L-1} e^{i\frac{q m \pi}{L}}e^{-i \frac{m \pi P}{L}} \mathcal{O} e^{i \frac{m\pi P}{L}}.
\label{eq:mqcdef}
\end{align}
The MQC intensities are defined as $I_q=2^{-L}\text{Tr}(\mathcal{O}_q \mathcal{O}_{-q})$. Using orthogonality condition and Eq.~(\ref{eq:mqcdef}) it can be shown that
\begin{align}
I_q=2^{-L}\text{Tr}(\mathcal{O}_q \mathcal{O}_{-q})=2^{-L}\text{Tr}(\mathcal{O}_q \mathcal{O})=\frac{2^{-L}}{2L}\sum_{m=0}^{2L-1} e^{i\frac{q m \pi}{L}}\text{Tr}(e^{-i \frac{m \pi P}{L}} \mathcal{O} e^{i \frac{m\pi P}{L}} \mathcal{O})=\frac{1}{2L}\sum_{m=0}^{2L-1} e^{i\frac{q m \pi}{L}} S_m
\label{eq:mqciq}
\end{align}
where $S_m=2^{-L}\text{Tr}(e^{-i \frac{m \pi P}{L}} \mathcal{O} e^{i \frac{m\pi P}{L}} \mathcal{O})$ is the signal we measure in the experiments (Eq.~(\ref{eq: mqc_sig}) in the main text). Measuring $S_m$ requires both forward and backward time evolution. Since $\mathcal{O}(t)=U_t\mathcal{O}(0)U_{-t}$, the signal at time $t$ can be written as 
\begin{align*}
S_m(t)=2^{-L}\text{Tr}(e^{-i \frac{m \pi P}{L}} U_t\mathcal{O}(0)U_{-t} e^{i \frac{m\pi P}{L}} U_t\mathcal{O}(0)U_{-t})=2^{-L}\text{Tr}(\mathcal{O}(0)U_{-t}e^{-i \frac{m \pi P}{L}} U_t\mathcal{O}(0)U_{-t} e^{i \frac{m\pi P}{L}} U_t)
\end{align*}
where we used the cyclic property of the trace in the second step. The MQC protocol consists of four steps: forward time evolution by $U_t$, rotation (phase tagging) by $e^{i \frac{m\pi P}{L}}$, backward time evolution by $U_{-t}$, and detection of initial state $\mathcal{O}(0)$. A schematic of the experimental protocol is illustrated in Fig. \ref{fig:mqcd}. 

Eq.~(\ref{eq:mqciq}) gives us a way to find $I_q$ experimentally without knowing the individual $\mathcal{O}_q$. 
One can use the second relationship in Eq. (\ref{eq:OOq}) and (\ref{eq:OqP}) to find $\mathcal{O}_q$ explicitly. Starting from $[P,\mathcal{O}]=\sum_q q \mathcal{O}_q$, by repeatedly commuting both sides with $P$ one can generate a linear system in $\mathcal{O}_q$:
\begin{align*}
[P,\mathcal{O}]&=\sum_q q \mathcal{O}_q \\
[P,[P,\mathcal{O}]]&=\sum_q q^2 \mathcal{O}_q \\
[P,[P,[P,\mathcal{O}]]]&=\sum_q q^3 \mathcal{O}_q \\
&\vdots \\
\underbrace{[P,[P,[P,[P,\cdots [P,}_{2L} \mathcal{O}],\cdots]]]]&=\sum_q q^{2L} \mathcal{O}_q.
\end{align*}
Using $\mathcal{O}_q^\pm=\mathcal{O}_q \pm \mathcal{O}_{-q}$, one can rewrite the above into two linear systems of equations:
\begin{align}
  \begin{bmatrix}
    [P,\mathcal{O}] \\
    [P,[P,[P,\mathcal{O}]]]\\
    [P,[P,[P,[P,[P,\mathcal{O}]]]]] \\
    \vdots
  \end{bmatrix}&=
    \begin{bmatrix}
    1 & 2 & 3 & \cdots \\
    1^3 & 2^3 & 3^3 & \cdots \\
    1^5 & 2^5 & 3^5 & \cdots \\    
    \vdots & \vdots & \vdots & \ddots 
  \end{bmatrix}
    \begin{bmatrix}
   \mathcal{O}^-_1 \\
    \mathcal{O}^-_2 \\
    \vdots \\
    \mathcal{O}^-_L 
  \end{bmatrix} \nonumber \\ 
    \label{eq:mqcfind} \\
  \begin{bmatrix}
    [P,[P,\mathcal{O}]] \\
    [P,[P,[P,[P,\mathcal{O}]]]]\\
    [P,[P,[P,[P,[P,[P,\mathcal{O}]]]]]] \\
    \vdots
  \end{bmatrix}&=
    \begin{bmatrix}
    1^2 & 2^2 & 3^2 & \cdots \\
    1^4 & 2^4 & 3^4 & \cdots \\
    1^6 & 2^6 & 3^6 & \cdots \\    
    \vdots & \vdots & \vdots & \ddots 
  \end{bmatrix}
    \begin{bmatrix}
   \mathcal{O}^+_1 \\
    \mathcal{O}^+_2 \\
    \vdots \\
    \mathcal{O}^+_L 
  \end{bmatrix} \nonumber
\end{align}
The constant matrices are in the form of Vandermondes matrix, and can be exactly inverted. Thus, by calculating nested commutators of $\mathcal{O}$ and $P$ one can find all the multiple quantum components $\mathcal{O}_q$. In practice we find Eq.~(\ref{eq:mqcfind}) to be more efficient at finding $\mathcal{O}_q$ than Eq.~(\ref{eq:mqcdef}). 

\subsection{OTO commutators involving the prethermal integrals of motion}
We are interested in measuring the OTO commutator $C_{\overline{Z}\mathcal{O'}}=-4L^{-1} \langle[\overline{Z}, \mathcal{O'}]^2\rangle_{\beta=0}$, where the prethermal integral of motion $\overline{Z}$ is given by the time average of operator $Z(t)$: $\overline{Z}=\frac{1}{t_\mathrm{pre}}\int_0^{t_\mathrm{pre}} Z(t)$, where $t_\mathrm{pre}$ is the timescale over which the prethermal conserved quantity is present. 
This commutator can be found by measuring the MQC intensities with different forward and backward evolution times. The measured signal in MQC experiments is given by
\begin{align*}
S_\phi(t_1,t_2)=2^{-L}\text{Tr}[ e^{-i \phi\mathcal{O'}}Z(t_1) e^{i \phi\mathcal{O'}}Z(t_2)]=\sum_q e^{-i q \phi}I_q(t_1,t_2)
\end{align*} 
where $t_1$ and $t_2$ are now independent. Integrate both sides with respect to $t_1$ and $t_2$ upto $t_\mathrm{pre}$ gives
\begin{align*}
2^{-L}\text{Tr}[ e^{-i \phi\mathcal{O'}}\overline{Z} e^{i \phi\mathcal{O'}}\overline{Z}]=\sum_q e^{-i q \phi}\frac{1}{t_\mathrm{pre}^2}\int_0^{t_\mathrm{pre}}dt_1\int_0^{t_\mathrm{pre}} dt_2 I_q(t_1,t_2) 
\end{align*}
expand both sides with respect to $\phi$ and equate terms in $\phi^2$ gives
\begin{align*}
C_{\overline{Z}\mathcal{O'}}=-4L^{-1}\langle [\overline{Z}, \mathcal{O'}]^2\rangle_{\beta=0}=\sum_q \frac{1}{t_\mathrm{pre}^2}\int_0^{t_\mathrm{pre}}dt_1\int_0^{t_\mathrm{pre}} dt_2 I_q(t_1,t_2) 
\end{align*}
In the experiments we can only measure $I_q(t_1,t_2)$ at discrete time points. If we replace the integrals by discrete sums we can measure
\begin{align*}
C_{\tilde{Z}\mathcal{O'}}=-4L^{-1}\langle [\tilde{Z}, \mathcal{O'}]^2\rangle_{\beta=0}=\sum_q\frac{q^2}{M^2}\sum_j^M \sum_k^M I_q(j \Delta t, k\Delta t)
\end{align*}
where $\tilde{Z}=\frac{1}{M}\sum_j Z(j \Delta t)$ is a discrete time average of $Z(t)$. The prethermal time scale $t_\mathrm{pre}$ is very large, so we assume the experimentally measured time is smaller than $t_\mathrm{pre}$. $\tilde{Z}$ can be used as an approximation for the integral of motion $\overline{Z}$.

\newpage
\section{IV. Constructing the Prethermal Hamiltonian}
\label{SMpre}
%

In this section we review the mathematical framework for prethermalization in time-independent systems. The derivations are based on Abanin et al.~\cite{Abanin2017}, but using a different approach to calculate the higher order correlations to the prethermal Hamiltonian. 
In prethermal systems, certain observables can exhibit slow, non-ergodic, dynamics even in the absence of any disorder. 
The reason  is that the prethermal Hamiltonian has an emergent symmetry given by $[H_\text{pre}, H_0']=0$ which is preserved for an exponentially long prethermal time. Thus, any operator that overlaps with $H_0'$ will have slow dynamics due to the emergent symmetry. The exponential timescale for $t_\text{pre}$ has been proved rigorously for both time-dependent and independent systems in~\cite{Abanin2017}.

To begin, it is useful to divide the transverse field dipolar Hamiltonian into two parts: $H_\text{TDip}= H_0 +V$ up to a scaling factor, where 
\begin{align}
H_0=\sum_j S_z^j, \qquad V= \epsilon\sum_{j,k>j}J_{jk}\!\left[S_y^j S_y^k -\frac{1}{2}(S_x^j S_x^k +S_z^j S_z^k) \right]
\end{align}
where $\epsilon$ is a small parameter. The idea is to perform a local unitary transformation to bring the Hamiltonian into the following form
\begin{align}
RHR^\dagger = H'_0 + D + \delta H = H_\text{pre} + \delta H
\label{eq:sw}
\end{align}
where $[D,H'_0]=0$, $[\delta H, H_\text{pre} ]\neq0$, and $H_\text{pre}=H_0' + D$. If $\delta H$ is zero, then $R^\dagger H_\text{pre} R$ is an exact conserved quantity. Any operators not orthogonal to $R^\dagger H_\text{pre} R$ will have some finite component conserved up to infinite times, and the prethermal time $t_\text{pre}$ is infinity. If, however, $\delta H$ is nonzero but exponentially small compared to $H_\text{pre}$, then $R^\dagger H_\text{pre} R$ is an emergent symmetry conserved only up to $t_\text{pre}$. 

In order to calculate the prethermal effective Hamiltonian $H_\text{pre}$, let $R=e^S$ and expand  $S$, and $D$ in powers of $\epsilon$:
\begin{align*}
S=\sum_{j=1} \epsilon^j S_j, \qquad D=\sum_{j=1} \epsilon^j D_j,
\end{align*}
Once $D_j$ are found the prethermal effective Hamiltonian up to $n_M$-th order is given by 
\begin{align}
H_\text{pre}=H'_0 + \sum_{j=1}^{n_M} \epsilon^j D_j
\label{eq:hpre}
\end{align}
It is not known whether the series in Eq.~(\ref{eq:hpre}) will converge for any $\epsilon$ when $n_M\rightarrow \infty$. When the series does converge then prethermal Hamiltonian is an exact symmetry. 

It is instructive to see how to solve this perturbation series iteratively. Using the Baker-Campbell-Hausdorff formula
\begin{align*}
e^S H e^{-S}=H + [S,H] + \frac{1}{2!}[S,[S,H]] + \frac{1}{3!}[S,[S,[S,H]]] + \frac{1}{4!}[S,[S,[S,[S,H]]]] +\cdots
\end{align*}
on the LHS of Eq.~(\ref{eq:sw}) and collecting terms with same order of $\epsilon$, it can be shown that the equations up to $\epsilon^5$ are 
\begingroup
\allowdisplaybreaks
\begin{align*}
D_1 + V'_1 &=[S_1, H_0] + V \\
D_2 + V'_2 &=[S_2, H_0] + [S_1, V] +\frac{1}{2} [S_1,[S_1,H_0]  \\
D_3 + V'_3 &=[S_3, H_0] + [S_2, V] +\frac{1}{2} ([S_1,[S_2,H_0]]+ 1\leftrightarrow 2) +\frac{1}{2} [S_1,[S_1,V]] \\ &+ \frac{1}{6} [S_1,[S_1,[S_1,H_0]]]\\
D_4 + V'_4 &=[S_4, H_0] + [S_3, V] +\frac{1}{2} ([S_1,[S_3,H_0]]+ 1\leftrightarrow 3) +\frac{1}{2} [S_2,[S_2,H_0]] \\ &+\frac{1}{2} ([S_1,[S_2,V]]+ 1\leftrightarrow 2)+ \frac{1}{6} ([S_2,[S_1,[S_1,H_0]]]+ 1\leftrightarrow 2) \\ &+ \frac{1}{6}[S_1,[S_1,[S_1,V]]] +\frac{1}{24}[S_1,[S_1,[S_1,[S_1,H_0]]]] \\
D_5 + V'_5 &=[S_5, H_0] + [S_4,V]+\frac{1}{2} ([S_1,[S_4,H_0]]+ 1\leftrightarrow 4)+\frac{1}{2} ([S_2,[S_3,H_0]]+ 2\leftrightarrow 3) \\ &+\frac{1}{2} ([S_1,[S_3,V]]+ 1\leftrightarrow 3) +\frac{1}{2} [S_2,[S_2,V]] +\frac{1}{6} ([S_2,[S_2,[S_1,H_0]]]+ 1\leftrightarrow 2) \\ &+ \frac{1}{6} ([S_3,[S_1,[S_1,H_0]]]+ 1\leftrightarrow 3) +\frac{1}{6} ([S_2,[S_1,[S_1,V]]]+ 1\leftrightarrow 2)\\ &+\frac{1}{24}([S_1,[S_1,[S_1,[S_2,H_0]]]]+1\leftrightarrow 2) +\frac{1}{24}[S_1,[S_1,[S_1,[S_1,V]]]] \\ &+ \frac{1}{120}[S_1,[S_1,[S_1,[S_1,[S_1,H_0]]]]]\\
&\vdots
\end{align*}
\endgroup
where $i\leftrightarrow j$ indicates permutation of $S_i$ and $S_k$ in the commutators. Starting at first order in $\epsilon$, all terms on the RHS besides $h_j\equiv[S_j, H_0]$  are split into two parts: the part that commutes with $H_0$ is stored in $D_j$; $S_j$ is then chosen such that $[S_j,H_0]$ exactly cancels the other part, which does not commute with $H_0$, thus automatically making $V'_j=0$. The iteration is repeated in the next order. 
\begin{figure}[b!]
\centering
\includegraphics[width=0.6\textwidth]{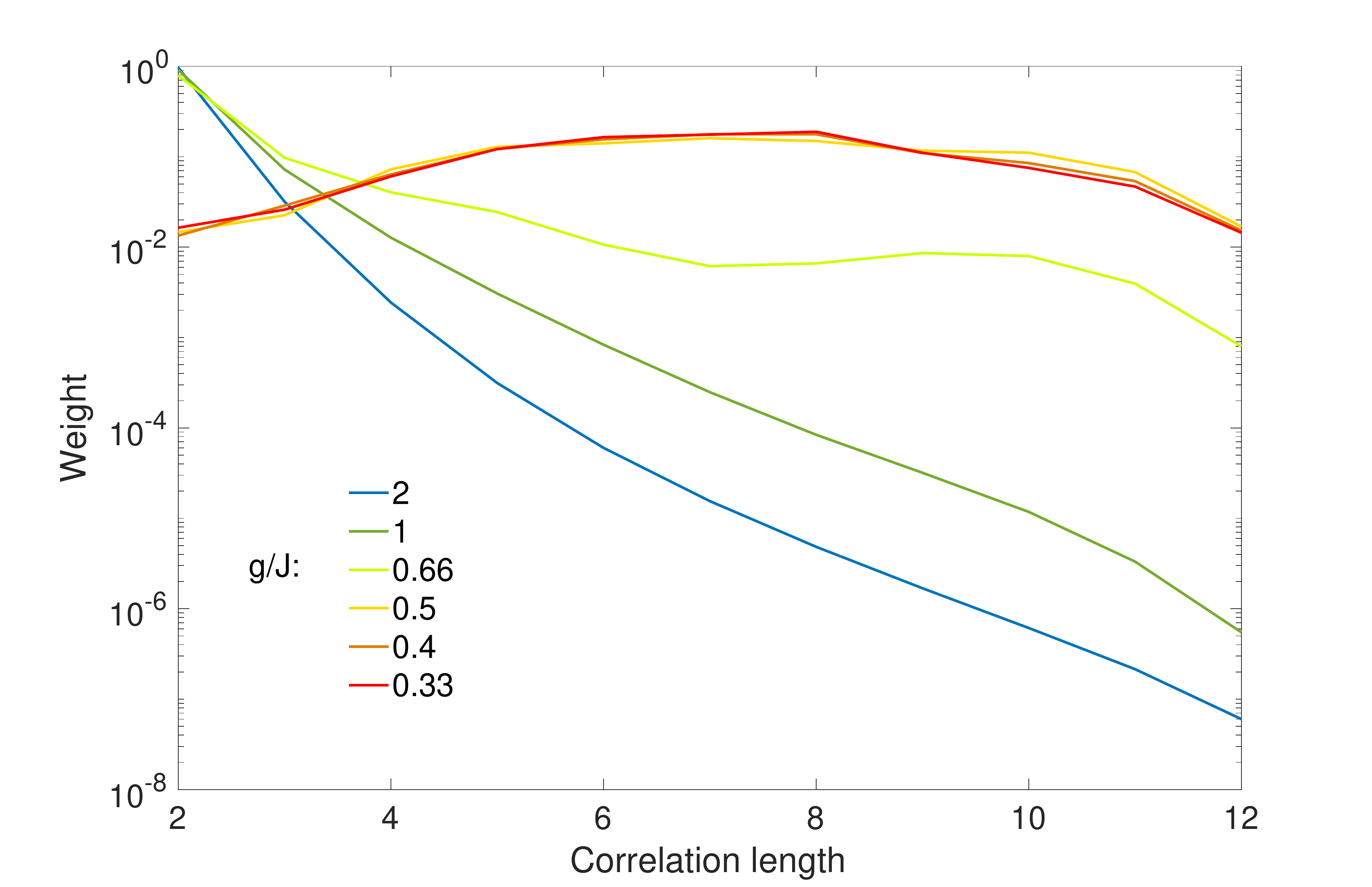}
\caption{\label{fig:locality}
\textbf{Locality of Prethermal Transformation.} We consider the unitary $R=e^S$ that transforms the transverse dipolar Hamiltonian into its prethermal form (see Eq.~(\ref{eq:hpre})). For a prethermal Hamiltonian, we expect  its generator $S$ to be local.  
Here we show the weight of terms with different correlation distances in $S$ at different values of $g/J$. $S$ is evaluated to 12th order in $L=12$ chain.
}
\end{figure}

Since $H_0$ satisfies $e^{2i m\pi H_0}=\mathbb{1}$, one can use the same approach for calculating MQC components to systematically and uniquely find $D_j$ and $S_j$. At each order, decompose $h_j$ to its MQC components: $h_j =\sum_q h_{jq}$, where $[H_0,h_{jq}]=q h_{jq}$, and find them using Eq.~(\ref{eq:mqcfind}). In terms of $h_{jq}$, $D_j$ and $S_j$ can be written as
\begin{align}
D_j = h_{j0},\qquad S_j =\sum_{q> 0} q^{-1}(h_{jq}-h_{j(-q)})
\end{align}
The MQC components $h_{jq}$ are referred to as generalized ladder operators in~\cite{Lin2017b}. If the original Hamiltonian contains only 2-body terms, $S_j$ contains at most $(j+1)$-body operator, guaranteeing the locality of the transformation. 

To confirm the locality numerically, we decompose $S$ to sum of products of Pauli operators. Each product operator is assigned a correlation distance defined as the distance between the furtherest non-identity operators (e.g. both $\sigma_0\otimes \sigma_1 \otimes\sigma_0 \otimes\sigma_2 \otimes\sigma_0$ and $\sigma_0\otimes \sigma_1 \otimes\sigma_3 \otimes\sigma_2 \otimes\sigma_0$ are of correlation distance 3). The weight of terms of different correlation distance is shown in Fig. \ref{fig:locality}. The local-to-nonlocal transition happens at around $g/J=0.5$. The problem of finding prethermal Hamiltonian is intimately related to finding the slowest operators, or the operators that best commutes with the Hamiltonian~\cite{Kim2015e, Lin2017b}.

\subsection{No Prethermalization for small $g$}
\begin{figure}[t!]
\centering
\includegraphics[width=90mm,clip]{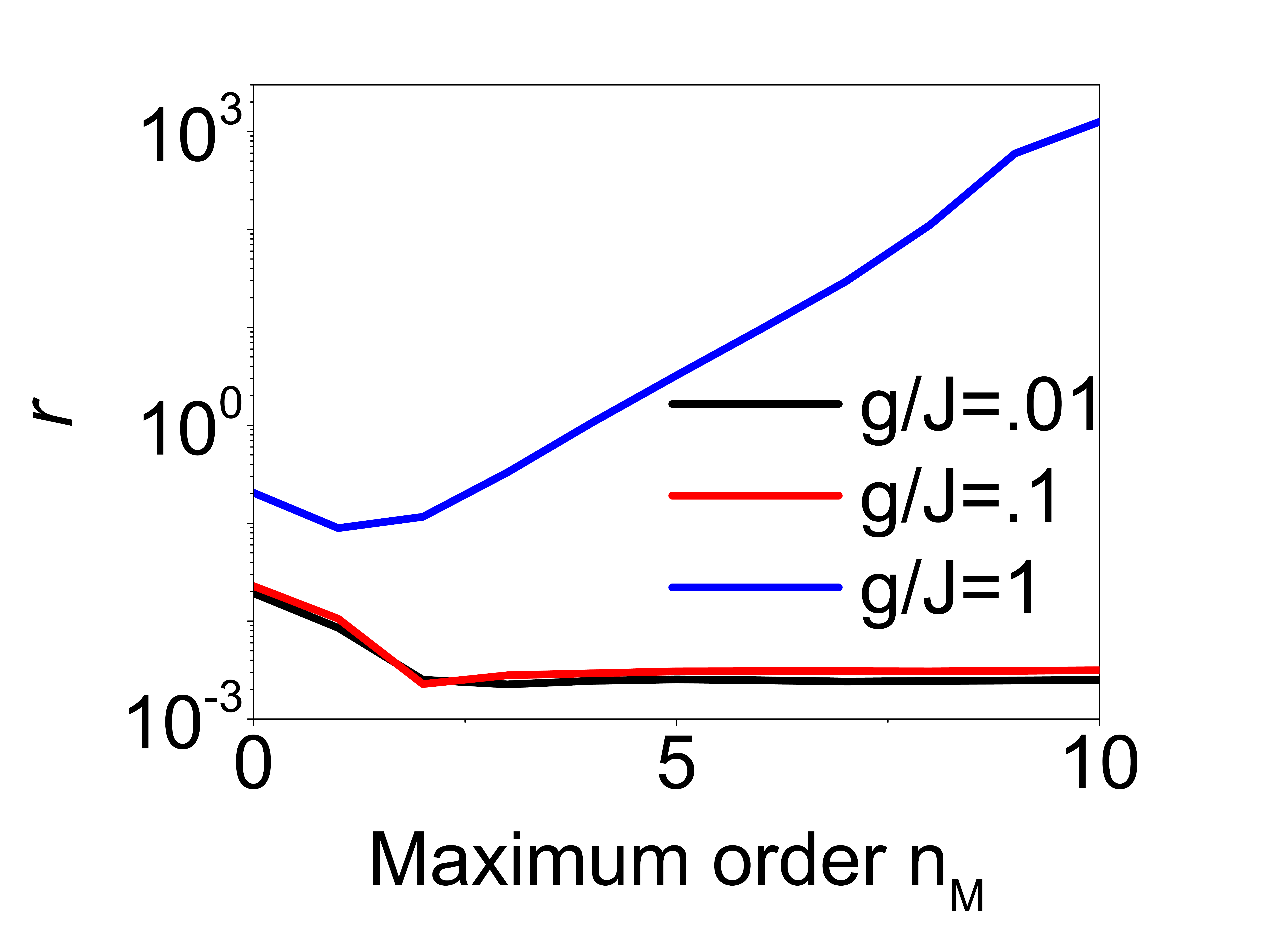}
\caption{\label{fig:preXX}
(a) Eigenvalue difference, $r$ with respect to maximum order, $n_M$, for varying values of $g/J$. 
$r$ fails to decrease after second order.
The system size is $L=11$.
}
\end{figure}
In Fig. \ref{fig:preXX} we explore whether the Hamiltonian in Eq. (\ref{eq:ham}) can be prethermal for small values of $g$. In the main text, it is shown that a prethermal Hamiltonian cannot be generated using $H_0=Z$ for small $g$. Another possible prethermal generator is $H_0=\sum_j 2S_y^jS_y^{j+1}$.  Following the method given in the previous section, we construct the prethermal Hamilontian order by order, and calculate the mean eigenvalue difference between the prethermal Hamiltonian and the original Hamiltonian. The eigenvalue difference $r$ stops decreasing after $n_M>2$ even for moderate transverse field strength (Fig. \ref{fig:preXX}), while for large transverse field it diverges as expected. The prethermalization theory fails to apply to small field regime because even with only dipolar interaction $||H_0||\approx ||V||$, thus the series expansion does not converge.

\section{V. Extended Data}

\subsection{Experimental data for $C_{YY}(t)$ and $C_{ZZ}(t)$}
\begin{figure}[h]
\centering
\includegraphics[width=0.8\textwidth]{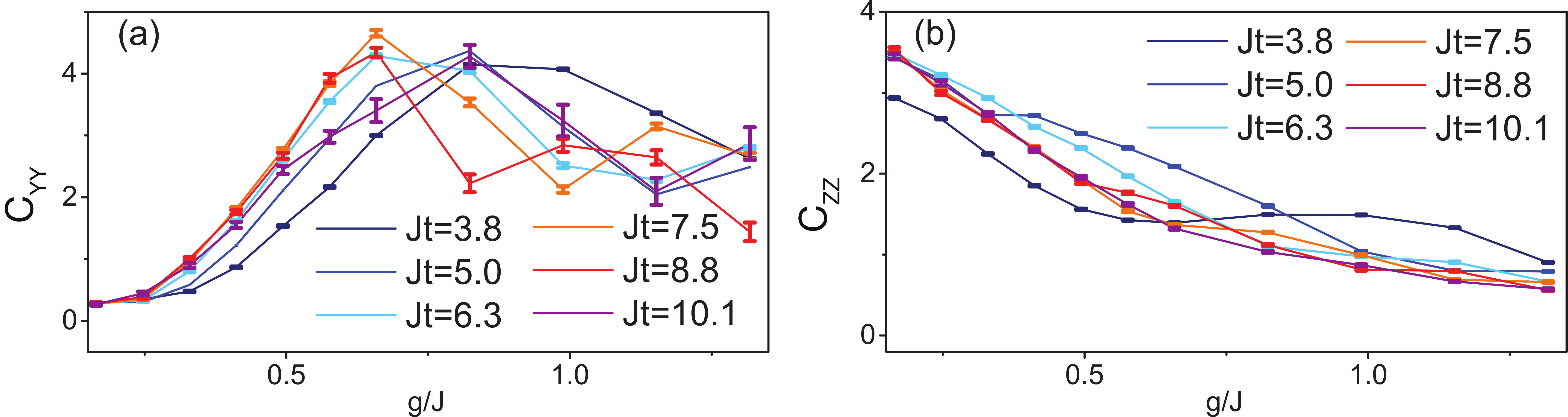}
\caption{\label{fig:XXZZfull}
Experimentally measured $C_{YY}(t)$ (a) and $C_{ZZ}(t)$ (b) versus transverse field at different times.}
\end{figure}

Figure \ref{fig:XXZZfull} shows the experimentally measured $C_{YY}(t)$ and $C_{ZZ}(t)$ at six different times, whose averaged value is presented in Fig. \ref{figCYZ}(c) in the main text.

\subsection{Experimental data for $\text{Tr}(Z(t)Z)$ and $\text{Tr}(Y(t)Y)$}
\begin{figure}[h]
\centering
\includegraphics[width=100mm,clip]{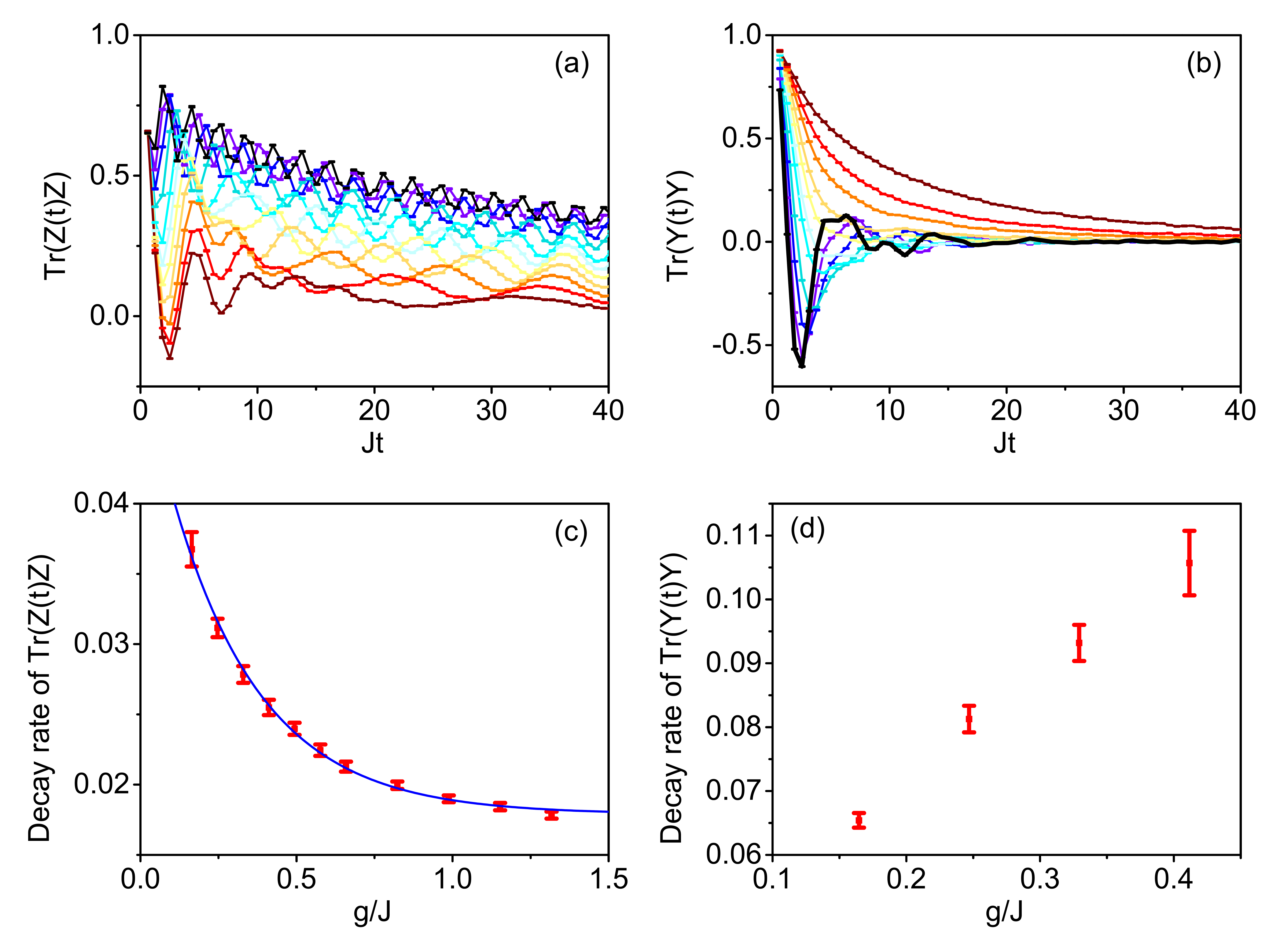}
\caption{\label{fig:mzmxfull}
Experimentally measured $\text{Tr}(Z(t)Z)$ (a) and $\text{Tr}(Y(t)Y)$ (b) versus time, with transverse field strength $g/J$=$\{0.16	, 0.25,	0.33,	0.41,	0.49,	0.58,	0.66,	0.82,	0.99,	1.2,	1.3\}$. Blue color represents larger field while red color represents smaller field. (c) Red markers show the decay rate of $\text{Tr}(Z(t)Z)$ as a function of field strength $J/g$. The error bar includes only fitting uncertainty, not including the uncertainty of the raw data. The decay rate can be fitted to an exponential function (blue curve). (d) Decay rate of $\text{Tr}(Y(t)Y)$ as a function of field strength $J/g$. For $J/g>0.41$, $\text{Tr}(Y(t)Y)$ cannot be fitted to an exponential function.}

\end{figure}

The extended data for $\text{Tr}(Z(t)Z)$ and $\text{Tr}(Y(t)Y)$ is shown in Fig. \ref{fig:mzmxfull}. The $\text{Tr}(Z(t)Z)$ exhibits oscillations on top of a decay. After averaging over one oscillation period, $\text{Tr}(Z(t)Z)$ can be fitted to an exponential function $A e^{-\gamma Jt}$, where $\gamma$ is the dimensionless decay rate. Figure \ref{fig:mzmxfull}(c) shows the decay rate $\gamma$ of $\text{Tr}(Z(t)Z)$ can be approximated by an exponential function of field strength $J/g$, i.e. $\gamma=\gamma_0 e^{-\alpha g/J}+\gamma_\infty$, where $\gamma_\infty$ is a background decay in the experiments that is independent of $g$. This exponentially slow decay agrees with the theory for Floquet prethermalization \cite{Else2017,Else2017a,Abanin2017}. $\text{Tr}(Y(t)Y)$ shows exponential decay at small $g/J$ [Fig. \ref{fig:mzmxfull}(b)], whose decay rate is shown in Fig. \ref{fig:mzmxfull}(d). The scaling for the decay rate of $\text{Tr}(Y(t)Y)$ cannot be extracted due to limited data points.

\newpage
\section{VI. Numerical results}
\begin{figure}[h]
\centering
\includegraphics[width=190mm,clip]{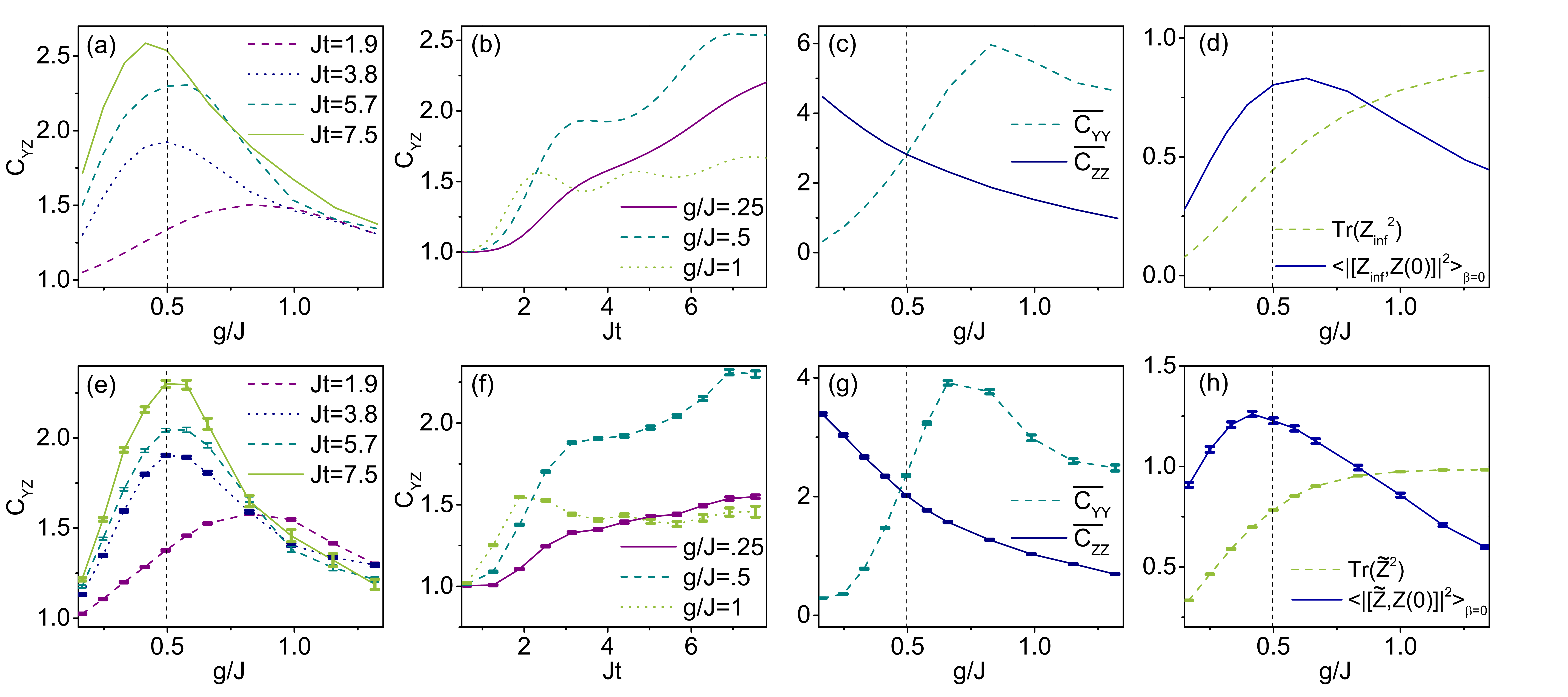}
\caption{\label{fig:otoc4_sim}
(a)  $C_{\mathrm{YZ}}$ with respect to transverse field strength for $Jt=1.9$ (purple dashed line), $Jt=3.8$ (dots), $Jt=5.7$ (green  dashed line) and $Jt=7.6$ (solid line).
(b) $C_{\mathrm{YZ}}$ as a function of normalized time, for $g=0.25$ (solid), $g=0.5$ (dashed) and $g=1$ (dots).
(c) Averaged $C_{\mathrm{YY}}$ (dashed) and $C_{\mathrm{ZZ}}$ (solid) with respect to transverse field strength. The time average is taken over the values $Jt=3.77, 5.02, 6.28, 7.54, 8.80, 10.05$.
(d)  $\mathrm{Tr}({Z}_\mathrm{Inf}^2)$ (dashed) and $\langle|[{Z}_\mathrm{Inf},Z(0)]|^2\rangle_{\beta=0}$ (solid) versus transverse field strength. 
The average is taken by keeping only the diagonal matrix elements of $Z$ in the eigenbasis of $H$.
(e)-(h) are the copy of Fig.~\ref{figCYZ} (b) and (c) and Fig. \ref{figAVE} for comparison.
}
\end{figure}

In this section we show the numerical results of OTO commutators using exact diagonalization (Fig. \ref{fig:otoc4_sim}). System size $L=12$ and open boundary condition are used here. In comparison with Fig. \ref{figCYZ} and Fig. \ref{figAVE} in the main text (reproduced here in (e)-(h) in Fig. \ref{fig:otoc4_sim}), the numerical and experimental results show quantitatively similar behavior. The difference is mainly due to experimental imperfections, decoherence and finite MQC encodings, as well as the finite size effect in simulation . In Fig. \ref{fig:otoc4_sim}(d), we calculated $Z_\text{inf}$ exactly by taking the diagonal ensemble in the eigenbasis: $Z_\text{inf}=\sum_n Z_{nn} \dyad{n}$, where $\ket{n}$ is an eigenvector of the Hamiltonian. The measured data in Fig. \ref{fig:otoc4_sim}(h) overestimates Fig. \ref{fig:otoc4_sim}(d) due to finite discrete averaging.

\bibliography{library1,Biblio}

\end{document}